\documentstyle[11pt,newpasp,twoside,epsf]{article}
\begin{document}
\title{Magnetic field topology of accreting white dwarfs}
 \author{Klaus Reinsch, Fabian Euchner, Klaus Beuermann}
\affil{Universit\"ats-Sternwarte, Geismarlandstr.\,11, 37083 G\"ottingen,
Germany}
\author{Stefan Jordan}
\affil{Institut f\"ur Astronomie und Astrophysik,
Eberhard-Karls-Universit\"at T\"ubingen, Sand 1,72076 T\"ubingen, Germany}

\begin{abstract}
We report first results of our systematic investigation of the magnetic field 
structure of rotating single magnetic white dwarfs and of white dwarfs in 
magnetic cataclysmic variables. The global magnetic field distributions on
the isolated white dwarf HE\,1045-0908 and the accreting white dwarfs in
EF\,Eri and CP\,Tuc have been derived from phase-resolved flux and polarization 
spectra obtained with FORS1 at the ESO VLT using the systematic method of 
Zeeman tomography. 
\end{abstract}

\section{Introduction}

Magnetic fields have been detected in $\sim$\,5\,\% of the 2300 known single
white dwarfs with field strengths ranging from $\sim$\,0.1--1000\,MG and 
peaking around 16\,MG (McCook \& Sion 1999, Wickramasinghe \& Ferrario 2000).
A similar fraction of magnetic white dwarfs has been found in accreting close 
binary systems (cataclysmic variables), possibly with a more restricted range 
of field strengths (7--230 MG, Beuermann 1998). The origin of the magnetic 
fields is not well understood. The magnetic fluxes, which should be conserved 
during stellar evolution, are similar to those of the magnetic main sequence 
stars such as the Ap and Bp stars which constitute $\sim$\,5\,\% of the 
normal main sequence stars. The decay times of the lowest multipole 
components are predicted to be long compared to the evolutionary ages of the 
white dwarfs. The magnetic field topologies, at least of isolated white dwarfs,
are, therefore, likely to be relics of previous evolutionary phases. In
accreting systems, the field structure in the outer layers of the white dwarf,
however, may have been significantly changed if the accretion rate is high
enough that accretion occurs more rapidly than ohmic diffusion (Cumming 2002).

Observational evidence suggests that the field topology of isolated and
accreting white dwarfs can deviate drastically from that of a centered dipole 
(e.g. Wickramasinghe \& Ferrario 2000, Schwope 1995). Information about the 
strength, orientation, and structure of the surface magnetic field of white 
dwarfs so far has been mainly derived from the analysis of photospheric Zeeman 
absorption lines and thermally broadened cyclotron harmonics from the polar 
regions of accreting white dwarfs. The interpretation of Zeeman and cyclotron 
intensity spectra alone is, however, often ambiguous and circular polarization 
spectra are required to investigate the magnetic field structure in detail. 

\section{Zeeman tomography}

We have developed a systematic method, called Zeeman tomography, to derive the 
global magnetic field distribution on rotating white dwarfs (Euchner et al.
2002). At present, it utilizes a database of 46800 sets of flux and circular 
polarization Zeeman spectra of homogeneous magnetic white dwarf atmospheres 
calculated with a code developed by S. Jordan for a broad range of field 
strengths $|\vec{B}| =$ 1--400\,MG, effective atmospheric temperatures $T =$ 
8000--50000\,K, and 9 field directions $\psi$ relative to the line of sight. 
All spectra are calculated for a surface gravity $\log g = 8$ and use a simple 
limb darkening law which is independent of wavelength. The magnetic field 
structure is approximated by a multipole expansion of the scalar magnetic 
potential, using spherical harmonics with coefficients $l = 1, \dots, 5$ and 
$m = 0$ for the zonal and sectoral periodicity of the field.
The database models are integrated for the $\vec{B}$ distribution of 900 
elements of the visible white dwarf surface at a given rotational phase. A 
least-squares optimization code based on an evolutionary strategy is used to 
reconstruct the multipole parameters from a set of flux and polarization 
spectra obtained at different rotational phases of the white dwarf. 

The Zeeman tomography method has been thoroughly tested with the reconstruction 
of various field geometries from synthetic spectra (Euchner et al. 2002). Here, 
we present first results of our application to real observational data.

\section{Observations}

We have obtained spin-phase resolved circular spectropolarimetry of a sample of
isolated and accreting magnetic white dwarfs using FORS1 at the ESO VLT during 
four observing runs between May 1999 and December 2000. The instrument has been
set up with a Wollaston prism and a rotatable quarter-wave retarder plate.
Exposures have been alternately taken with retarder plate position angles $\phi
= -45\deg$ and $\phi = +45\deg$ in order to correct for instrumental
polarization effects and linear-polarization cross-talk during the data 
reduction. Spectra of the target star and comparison stars in the field 
have been obtained simultaneously by using the multi-object spectroscopy mode 
of FORS. This allowed us to derive individual correction functions for the
atmospheric absorption losses in the target spectra and to check the data for 
remnant instrumental polarization. The target spectra cover the wavelength 
range $3800-7900$\,\AA\ at a spectral resolution $\lambda/\Delta\lambda = 440$. 
A signal-to-noise ratio $S/N \sim 100$ has been reached for the individual 
flux spectra at a spin-phase resolution $\Delta\Phi \sim 0.2$. All raw 
data have been reduced using the context MOS of the ESO MIDAS package.

All our spectropolarimetric observations of magnetic CVs have been obtained
while the systems were in a low-state of accretion. Hence, the Zeeman-split 
photospheric hydrogen absorption lines were clearly visible as the emission 
of the white dwarf was not diluted by emission from the accretion stream.

\begin{figure}
\plotone{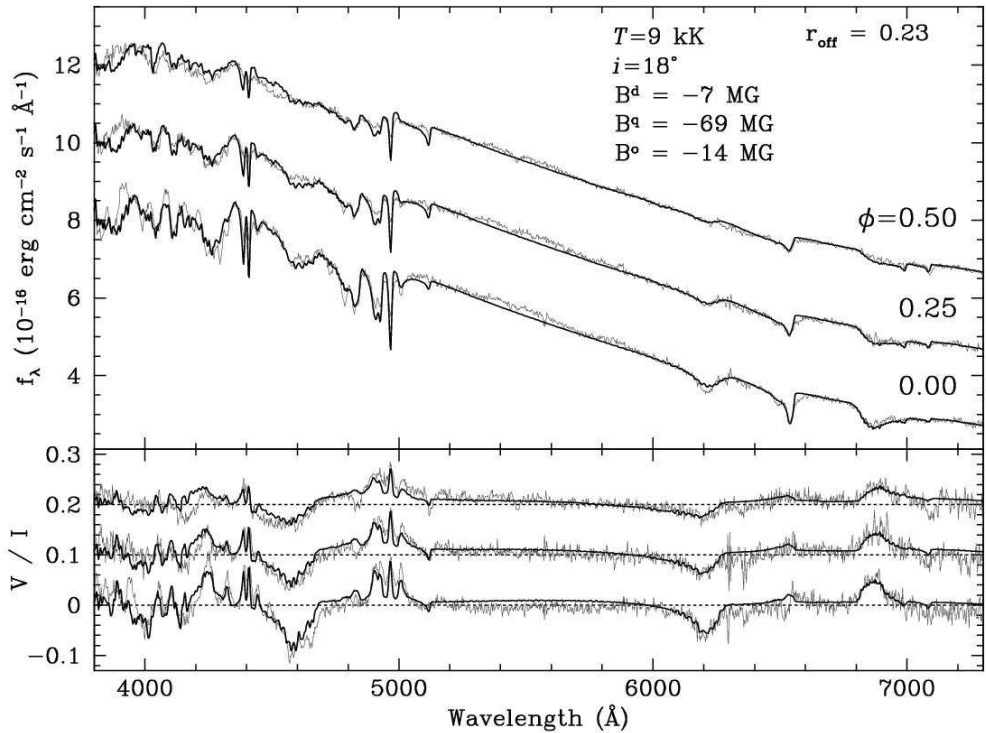}
\caption{Flux {\it (top)} and circular polarization {\it (bottom)} spectra of 
HE\,1045-0908 at (the arbitrary) rotational phases $\Phi =$ 0.0, 0.25, and 0.5. 
The synthetic spectra for the best-fit model {\it (thick line)} consisting of 
a dipole, quadrupole, and octopole field component is shown superimposed on the 
observed spectra {\it (grey curve)}. For clarity, the upper two flux spectra 
have been offset by 2 and 4 flux units, respectively, and the polarization
spectra by 0.1 and 0.2 units, respectively.}
\end{figure}

\begin{figure}
\plottwo{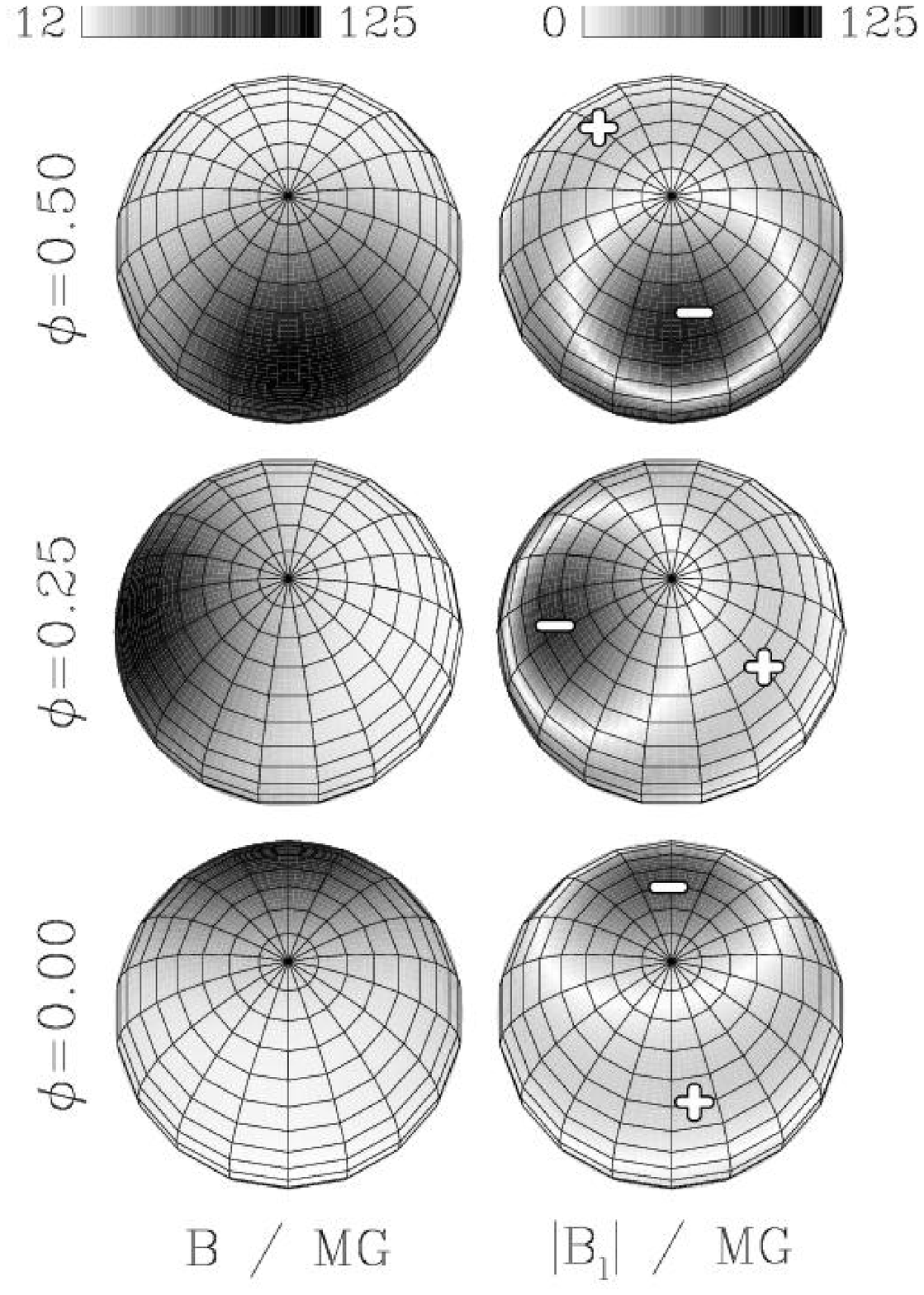}{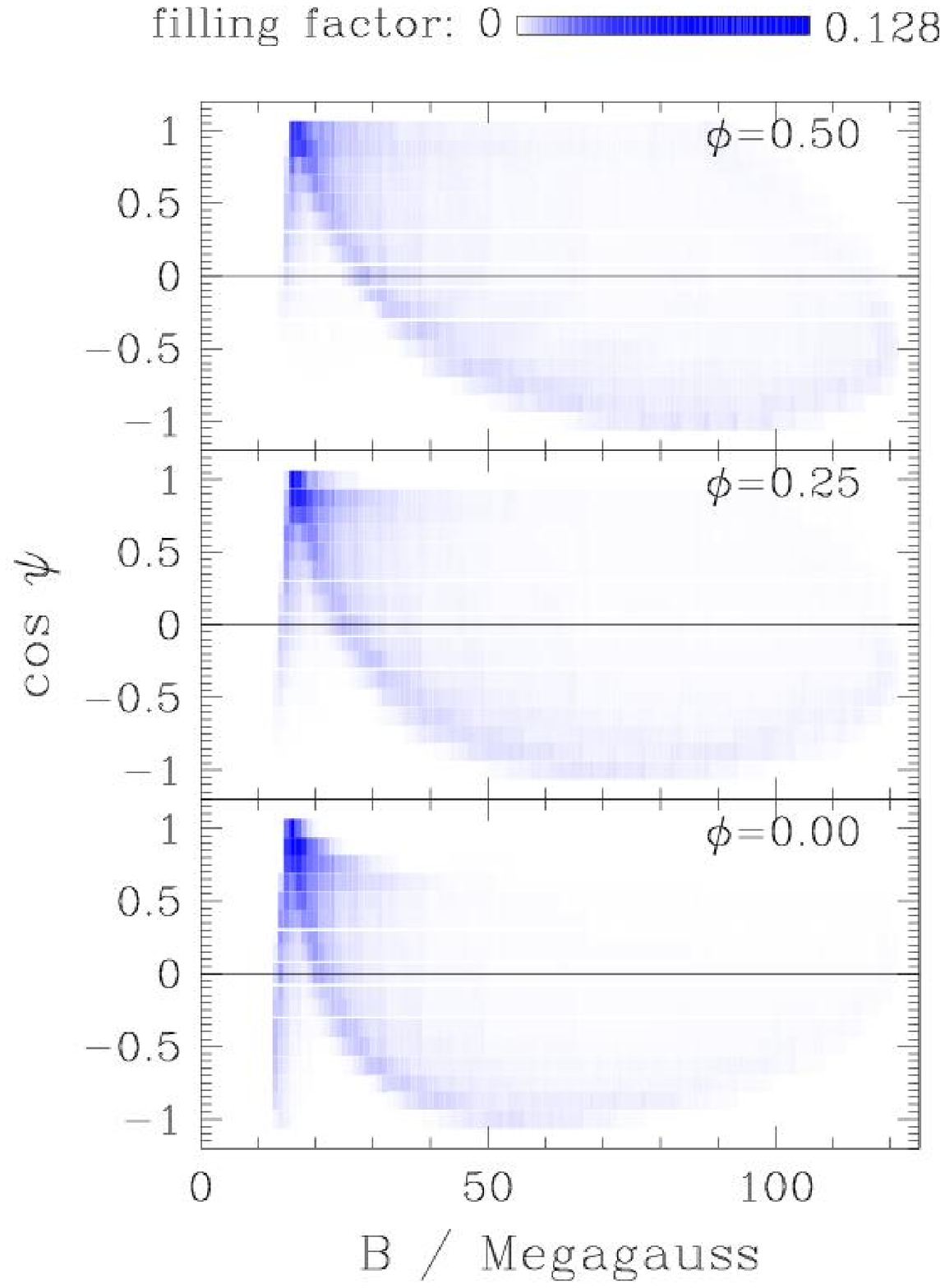}
\caption{Distribution of the total magnetic field strength $B$ and its 
longitudinal component $B_{\rm l}$ on the white dwarf in HE\,1045-0908 at 
(arbitrary) rotational phases $\Phi = $0.0, 0.25, and 0.5 {\it (left)} and 
frequency distribution of $B$ and the viewing direction cosine $\cos \Phi$ 
{\it (right)}.}
\end{figure}

\begin{figure}
\plotone{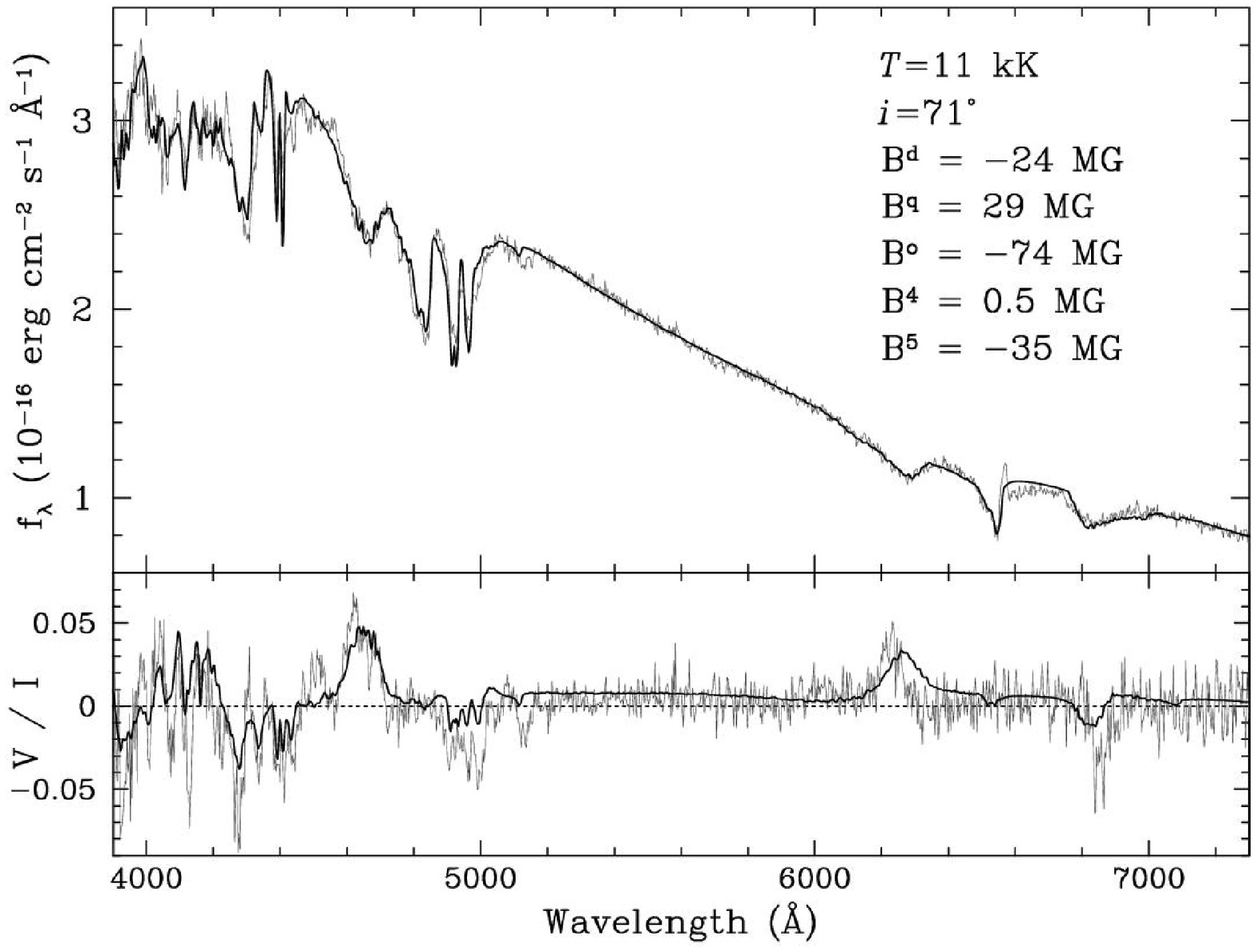}
\caption{Average flux {\it (top)} and circular polarization {\it (bottom)} 
spectrum of EF Eri. 
The synthetic spectrum for the best-fit model {\it (thick line)} consisting of 
a five component multipole expansion is shown superimposed on the observed 
spectrum {\it (grey curve)}.}
\end{figure}

\section{Results}

\subsection{HE\,1045-0908}

HE\,1045-0908 is an isolated hydrogen-line white dwarf with a polar field 
strength $B \sim 20$\,MG and a rotational period $P_{\rm rot} \sim 2-4$\,hr 
(Reimers et al. 1994, Schmidt et al. 2001).
Our observations cover $\sim$ 2\,hr during which drastic variability in the 
flux and polarization spectra has been seen (Fig. 1).
The field topology can be described by a combination of aligned dipole,
quadrupole, and octopole terms with a common offset perpendicular to the
magnetic axis (Fig. 2). 
An almost equally well fit has been obtained for a multipole expansion 
including the zonal components up to $l = 5$. It is reassuring that both fits 
yield similar ranges for the observed field strengths and viewing angles.

\subsection{EF\,Eri}

Spectropolarimetric observations of EF\,Eri have been obtained during two
epochs, in December 1999 and in November 2000 and cover in total 2 spin cycles
of the white dwarf. 
Rotational phases have been calculated using the ephemeris given by Piirola et 
al. (1987), with $\Phi = 0.0$ corresponding to the narrow minimum in the infrared
light curve of EF Eri, i.e. to the phase when the magnetic pole is closest to 
the observer.
As our data show little variation in the Zeeman features modeling has been 
done for the phase averaged flux and polarization spectra (Fig. 3 and 4). 
The data are sufficiently well described by a multipole expansion including 
the zonal components up to $l = 5$. We note, however, that there are still
systematic residuals in our best fit solution which indicate that the field
topology must be even more complex.

\begin{figure}
\plottwo{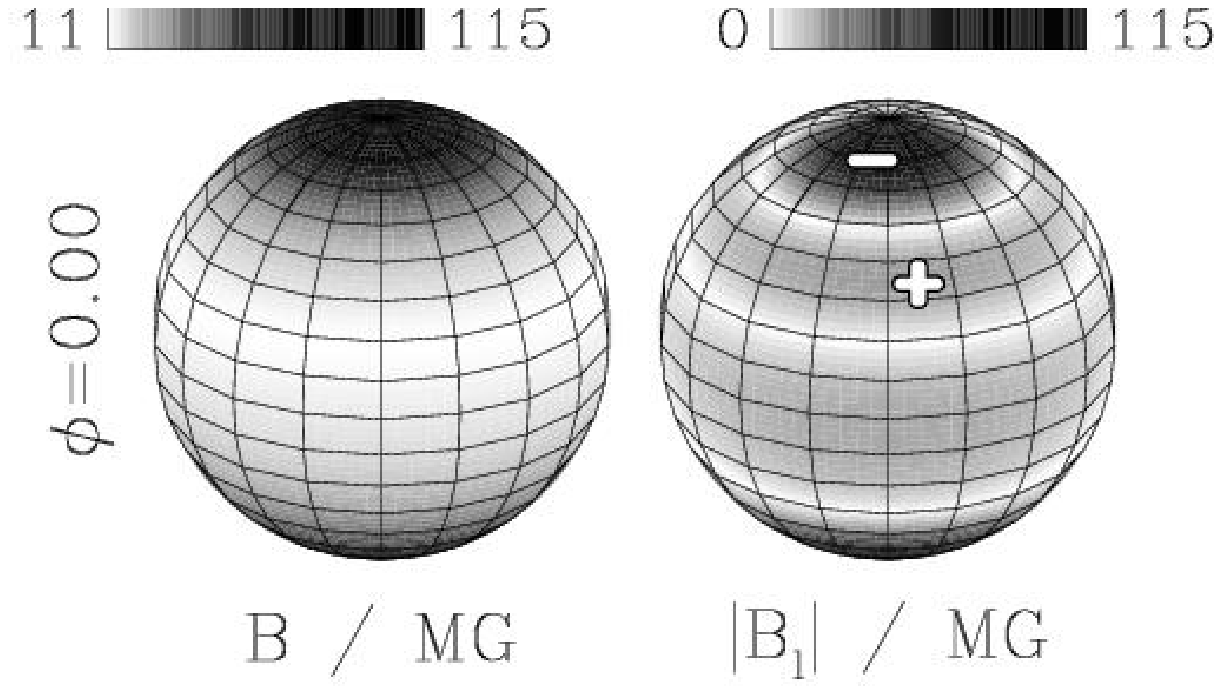}{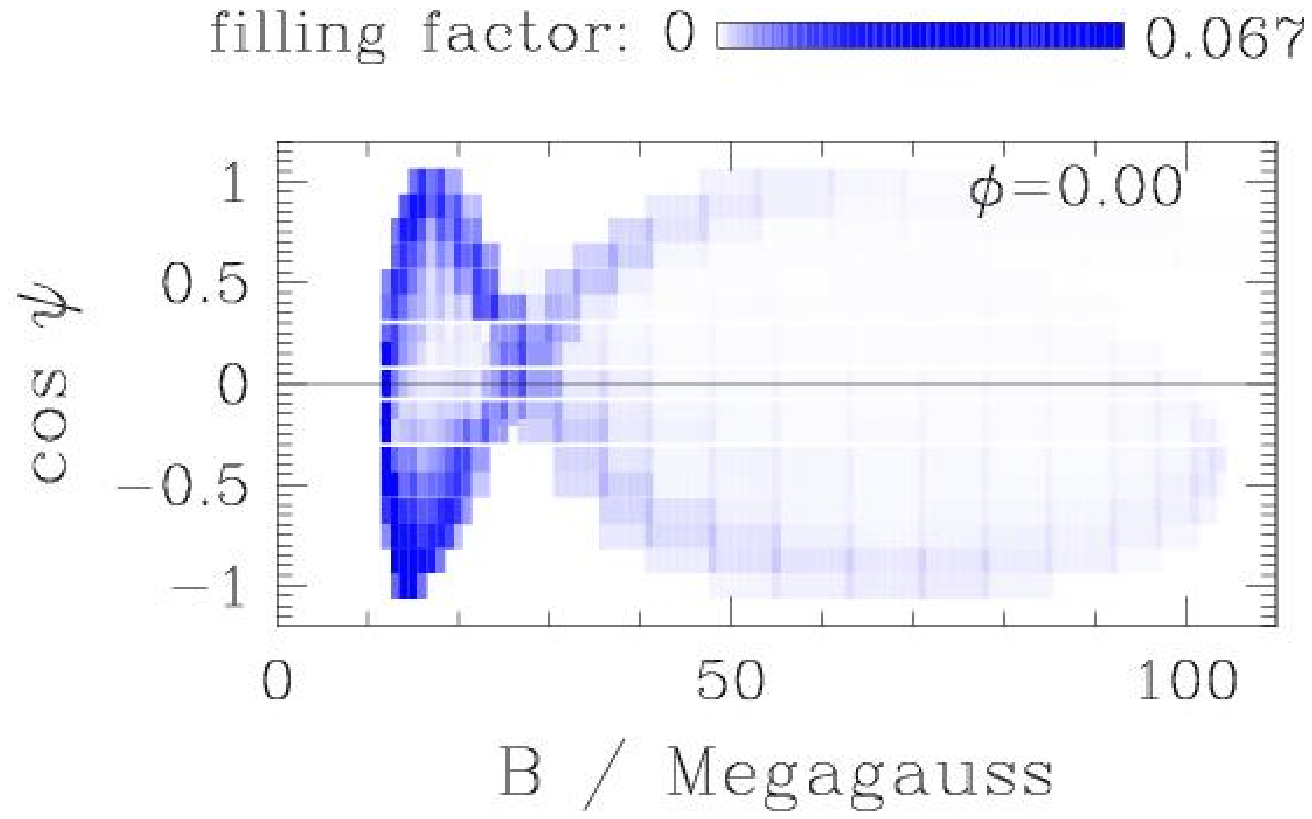}
\caption{Distribution of the total magnetic field strength $B$ and its 
longitudinal component $B_{\rm l}$ on the white dwarf in EF\,Eri at spin 
phase $\Phi = 0.0$ {\it (left)} and frequency distribution of $B$ and the 
viewing direction cosine $\cos \Phi$ {\it (right)}.}
\end{figure}

\subsection{CP\,Tuc}

CP\,Tuc (= AX\,J2315-592) has been discovered by Thomas \& Reinsch (1996) who
derived an upper limit of the magnetic field strength at the accreting pole $B
\le 17$\,MG from the properties of the optically thin cyclotron emission.
Our spectropolarimetric observations cover $\sim $0.5 spin periods of CP\,Tuc
(Fig. 5 and 6). 
Rotational phases have been calculated using the ephemeris given by Ramsay et 
al. (1999) where $\Phi = 0.0$ is defined by the minimum of the X-ray light curve.
Our best-fit solution for the field topology comprises a dipole, quadrupole 
plus octopole combination. Again, residual features indicate an even more
complex field. The high polar field strength found here is not contradicting 
the much lower upper limit derived by Thomas \& Reinsch (1996) for the field 
strength in the accretion region as the latter is probably offset from the 
magnetic poles.

\begin{figure}
\plotone{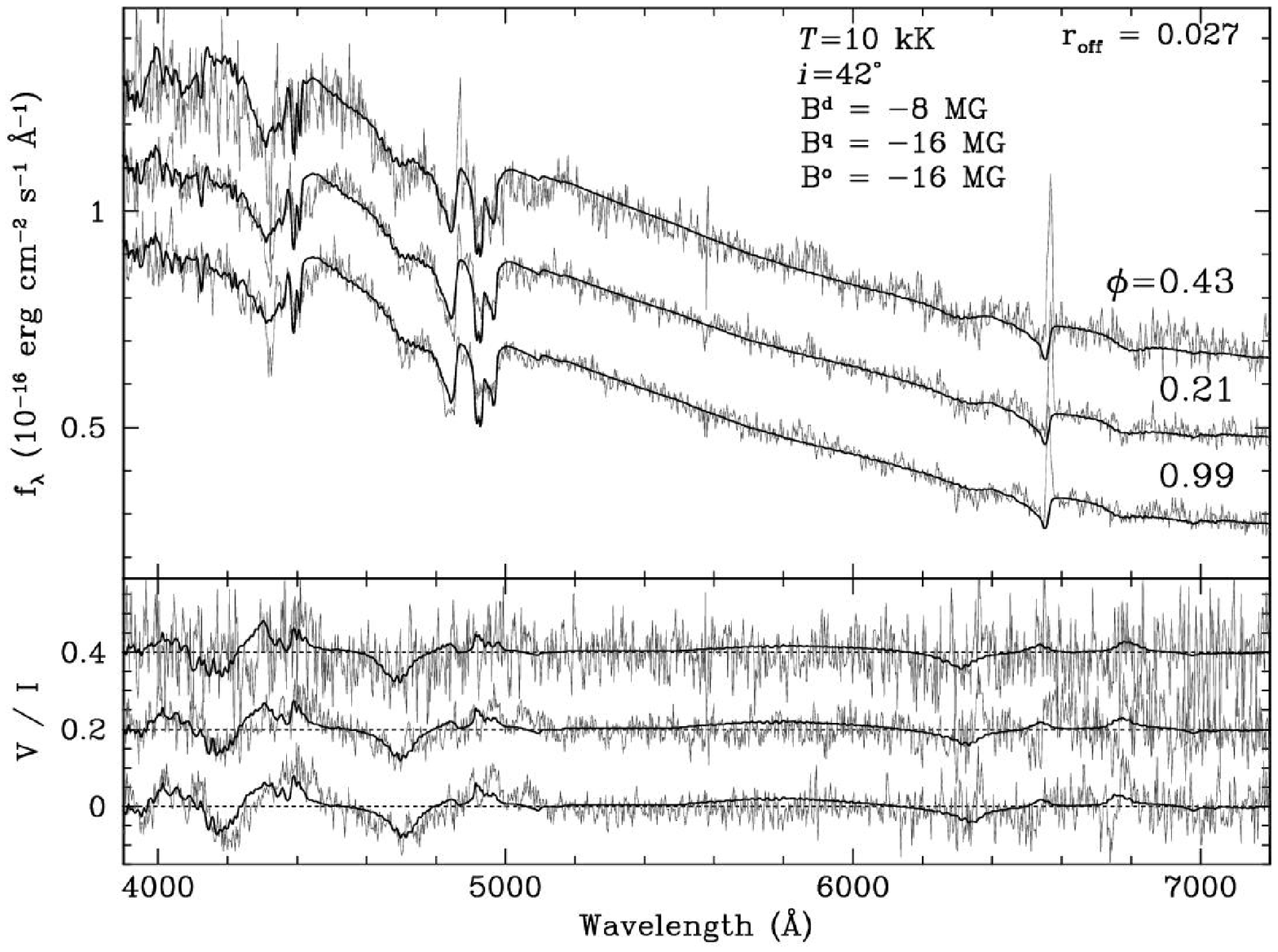}
\caption{Flux {\it (top)} and circular polarization {\it (bottom)} spectra of 
CP\,Tuc at spin phases $\Phi =$ 0.99, 0.21, and 0.43. 
The synthetic spectra for the best-fit model {\it (thick line)} consisting of 
a dipole, quadrupole, and octopole field component is shown superimposed on the 
observed spectra {\it (grey curve)}. 
The upper two flux and polarization spectra have been offset by 0.2 and 0.4 
units, respectively.}
\end{figure}

\begin{figure}
\plottwo{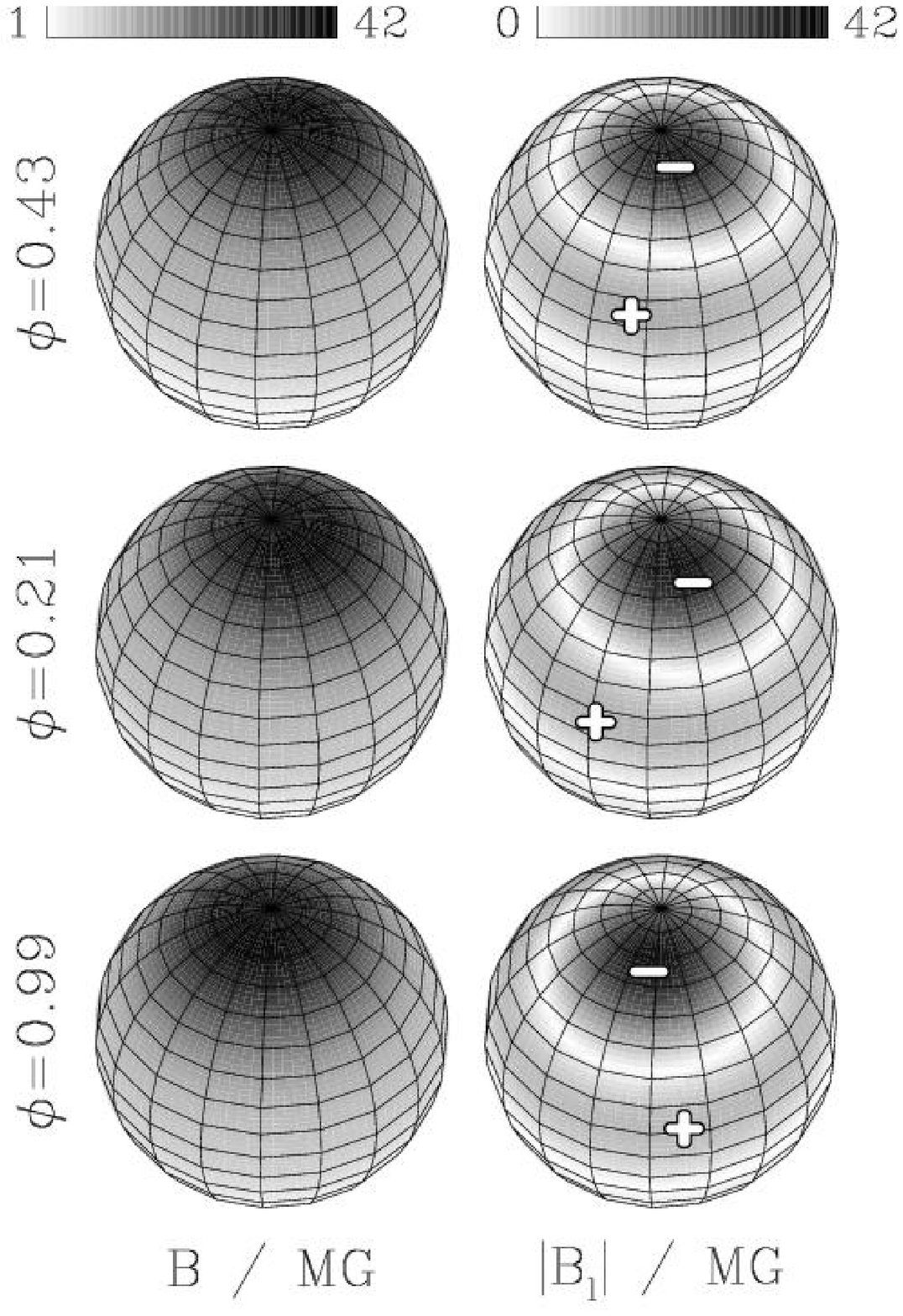}{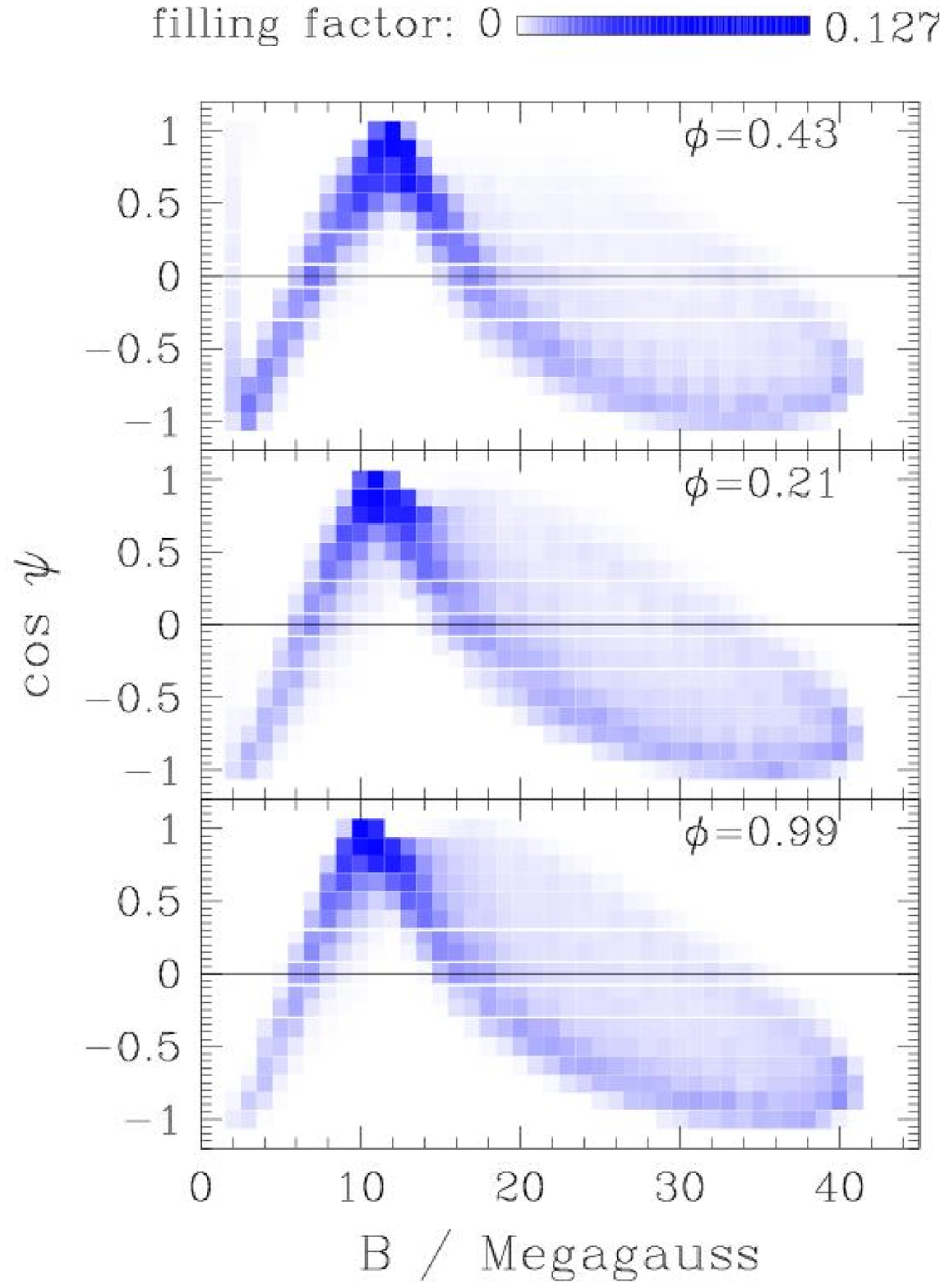}
\caption{Distribution of the total magnetic field strength $B$ and its 
longitudinal component $B_{\rm l}$ on the white dwarf in CP\,Tuc at rotational 
phases $\Phi = $0.99, 0.21, and 0.43 {\it (left)} and frequency distribution 
of $B$ and the viewing direction cosine $\cos \Phi$ {\it (right)}.}
\end{figure}

\section{Conclusions}

We have shown that Zeeman tomography is a suitable systematic method to
derive the global magnetic field distribution on rotating white dwarfs.
With this method we have obtained for the first time detailed information 
about the range of field strengths and the field topology of a sample of
isolated and accreting white dwarfs. Our results clearly demonstrate that a 
single value as obtained with hitherto available methods is not sufficient
to quantify the field of magnetic white dwarfs. Overall our model
fits are in excellent agreement with observations. Remaining differences 
indicate that the field topology is even more complex than described by an
up to 5 component multipole expansion. A more detailed discussion of our 
results will be presented elsewhere (Euchner et al., in prep.).

\acknowledgements
Based on observations collected at the European Southern Observatory, Chile
under program numbers 63.P-0003(A), 64.P-0150(C), and 66.D-0128(A).
This work was supported in part by BMBF/DLR grant 50\,OR\,9903\,6.


\begin{references}
\reference Beuermann, K. 1998, in {\it High Energy Astronomy and Astrophysics},
  Tata Inst. of Fund. Res., 100
\reference Cumming, A. 2002, \mnras, 333, 589
\reference Euchner, F., Jordan, S., Beuermann, K., G\"ansicke, B.T., 
  Hessman, F.V. 2002, \aap, 390, 633
\reference McCook, G.P., Sion, E.M. 1999, \apjs, 121, 1
\reference Piirola, V., Coye, G.V., Reiz, A. 1987, \aap 186, 120
\reference Ramsay, G., Potter, S.B., Buckley, D.A.H., Wheatley, P.J. 1999, 
  \mnras, 306, 809
\reference Reimers, D., Jordan, S., K\"ohler, T., Wisotzki, L. 1994, 
  \aap, 285, 995
\reference Schmidt, G.D., Vennes, S., Wickramasinghe, D.T., Ferrario, L. 2001,
  \mnras, 328, 203
\reference Schwope, A.D. 1995, Rev. Mod. Astron., 8, 125
\reference Thomas, H.-C., Reinsch, K. 1996, \aap, 315, L1
\reference Wickramasinghe, D.T., Ferrario, L. 2000, \pasp, 112, 873
\end{references}
\end{document}